\documentclass{aa}

\usepackage{natbib,twoopt}
\usepackage[breaklinks=true]{hyperref} %% to avoid \citeads line fills
\bibpunct{(}{)}{;}{a}{}{,}             %% natbib format for A&A and ApJ
\usepackage{graphicx}
\usepackage{txfonts}
\usepackage{hyperref}
\hypersetup{
    colorlinks=true,
    linkcolor=blue,
    citecolor = blue,
    filecolor=magenta,      
    urlcolor=blue}
\newcommand{\angstrom}{\mbox{\normalfont\AA}}
\graphicspath{{./}{Figures/}}

\authorrunning{Alvarez Garay et al.}
\titlerunning{Mg-K anticorrelation in NGC1786}

\def\fe{$\rm [Fe/H]$}

\begin{document}

\title{The chemical DNA of the Magellanic Clouds}
\subtitle{III. The first, extragalactic Mg-K anticorrelation: the LMC globular cluster NGC~1786}
%\title{First Mg-K anticorrelation in an extragalactic globular cluster}

\author{Deimer Antonio Alvarez Garay\inst{1},
Alessio Mucciarelli\inst{2}\fnmsep\inst{3},
\and Lorenzo Monaco \inst{4}\fnmsep\inst{5}
}

   \institute{Osservatorio Astrofisico di Arcetri, Largo E. Fermi 5, 50125
              Firenze, Italy\\
              \email{deimer.alvarez@inaf.it}
              \and
              Dipartimento di Fisica e Astronomia, Università degli Studi di Bologna, Via Gobetti 93/2, I-40129 Bologna, Italy
              \and
              INAF, Osservatorio di Astrofisica e Scienza dello Spazio di Bologna, Via Gobetti 93/3, I-40129 Bologna, Italy
              \and
              Universidad Andres Bello, Facultad de Ciencias Exactas, Departamento de F{\'i}sica y Astronom{\'i}a - Instituto de Astrof{\'i}sica, Autopista Concepci\'on-Talcahuano 7100, Talcahuano, Chile
              \and
              INAF-OATs, Via G.B.Tiepolo 11, Trieste, I 34143, Italy
              }

%   \date{Received September 15, 1996; accepted March 16, 1997}

% \abstract{}{}{}{}{} 
% 5 {} token are mandatory
 
\abstract
{In this work we derived [K/Fe] and [Mg/Fe] abundance ratios for six stars of the old globular cluster NGC~1786 
in the Large Magellanic Cloud.  We employed high-resolution spectra acquired with the MIKE spectrograph mounted at the Magellan/Clay telescope. We found a clear Mg-K anticorrelation among the analyzed stars. In particular, the Mg-poor stars ([Mg/Fe] $< 0.0$ dex) are enriched by $\sim 0.25$ dex in [K/Fe] compared to the Mg-rich stars ([Mg/Fe] $> 0.0$ dex). This finding makes NGC~1786 the first globular cluster 
residing in an external galaxy in which such extreme chemical anomaly has been detected. 
The observed trend nicely agrees with those observed in Galactic globular clusters hosting Mg-poor stars, such as NGC~2808, and $\omega$ Centauri suggesting that such chemical anomaly is an ubiquitous feature of old, massive, and metal-poor stellar systems and it does not depend on the properties of the parent galaxy in which the cluster formed. Also, Na-O and Mg-Al anticorrelations were detected among the stars of NGC~1786. The newly discovered Mg-K anticorrelation reinforces the idea that stars capable of activating the complete MgAl cycle are responsible for the observed chemical anomalies in these clusters. In this context, asymptotic giant branch stars seem to be a valuable model since they are able to produce K while depleting Mg. However, the precise and complete physics of this model remains a subject of debate.
}

\keywords{globular clusters: individual (NGC 1786) – stars: abundances – techniques: spectroscopic}
   \maketitle
%
%-------------------------------------------------------------------

\section{Introduction}
In the last decade, potassium (K) has earned an important place in the restricted group of key chemical elements for characterizing the chemical evolution of globular clusters (GCs). The first evidence of an intrinsic K spread was discovered in the massive and metal-poor GC NGC 2419, where \citet{mucciarelli_12} and \citet{cohen_12} independently identified a clearcut Mg-K anticorrelation. 
In this cluster, a subset of stars exhibits extreme magnesium (Mg) depletion ([Mg/Fe] down to $\sim -1$ dex) coupled with an extreme enhancement of K ([K/Fe] up to $\sim +2$  dex). This feature, which is exceptionally pronounced in NGC~2419, was identified in only a handful of other GCs belonging to the Milky Way, namely NGC~2808 \citep{mucciarelli_15}, NGC~4833 \citep{carretta_21}, 
M~54 \citep{carretta_22}, and $\omega$ Centauri \citep{alvarezgaray_22}. 
All these systems are among the most massive and/or metal-poor in the Galaxy and some of them are distinguished by the presence of a sub-population of Mg-poor stars ([Mg/Fe] $< 0.0$ dex). 
Other globular clusters \citep[i.e. NGC~104, NGC~6752 and NGC~6809,][]{mucciarelli_17} without a Mg-poor sub-population 
show small or null [K/Fe] spreads, pointing out that the K enhancement is found only in Mg-poor stars.
The presence of such chemical signature is interpreted as the result of an extreme self-enrichment process. 

Indeed, the chemical inhomogeneities observed in GCs, involving (anti)correlations among light elements such as C, N, O, Na, Mg, Al, Si, and K, are widely interpreted as the signature of this self-enrichment history, where multiple populations of stars were born within the clusters \citep{bastian_18, gratton_19, milone_22}. 
The majority of theoretical models devoted at describing the formation and evolution of GCs posit that GCs experienced multiple episodes of star formation within the first 100–200 Myr of their life.
In this scenario, a second population (2P) of stars was formed from gas enriched by a first population (1P) of massive stars. These 1P stars, born with a field-like chemical composition, processed material through the hot CNO cycle and its secondary NeNa and MgAl chains \citep[e.g.;][]{langer_93,prantzos_07}. This chemically altered material was then ejected at low velocity into the intracluster medium, from which the 2P stars were born.

Different polluters were proposed in the literature to explain the presence of different populations of stars in GCs \citep[see e.g.;][for a full discussion]{renzini_15,bastian_18,milone_22}. However, these self-enrichment models fail to reproduce all the chemical patterns observed so far. In particular, the Mg-K anticorrelation is an important evidence for the theoretical models, since the presence of such chemical feature points to very high-temperature hydrogen burning (T $> 10^8$ K) reactions, where proton captures on argon (Ar) nuclei can synthesize K \citep{ventura_12, iliadis_16,prantzos_17}. Therefore, as a result, the analysis of Mg and K can place substantial constraints on the nature of polluters responsible for the chemistry seen in GCs. 
%However, up to now the Mg-K anticorrelation was detected only in GCs belonging to the Milky Way. We do not know if the manifestation of this extreme chemical feature is influenced by external parameters such as the environment of GCs formation, or if it depends only on the properties of GCs (mass and metallicity mainly).

While key chemical signatures of self-enrichment, such as the Na-O and Mg-Al anticorrelations, were confirmed to be present in extragalactic GCs \citep{mucciarelli_09}, suggesting they are a ubiquitous feature of old, massive stellar systems regardless of the parent galaxy, the Mg-K anticorrelation was observed to date only in a few massive, Milky Way clusters. This raises a fundamental question: is the manifestation of this extreme chemical feature influenced by the formation environment of the host galaxy, or does it depend solely on intrinsic GC properties, primarily mass and metallicity? The GC NGC~1786 represents an ideal benchmark to challenge this question. Indeed, this cluster belongs to the Large Magellanic Cloud (LMC), it is old and coeval to the Milky Way clusters \citep{brocato96}, metal-poor  \citep[\fe=--1.72 dex;][]{mucciarelli_21}, 
massive \citep[$\rm 5\times10^{5}~M_{\odot}$;][]{mackey2003} and harboring some, rare Mg-poor stars \citep{mucciarelli_09}.
It provides the unique opportunity to check the existence of a Mg-K anticorrelation in 
a genuine extragalactic GC where Mg-poor stars exist. \\
This paper is organized as follows: In Sect. \ref{sect_obs} we will present the dataset; in Sect. \ref{sect_analysis} we will describe the method we adopted to derive chemical abundances, while in Sect. \ref{sect_res} we will present the results of our chemical analysis. Finally, in Sect. \ref{sect_disc} we discuss and summarize our findings.

%-------------------------------------------------------------------
\begin{table*}
\caption{Information about the sample stars of NGC~1786. Coordinates are from \citet{gaia_16,gaia_23} and atmospheric parameters from \citet{mucciarelli_21}.}
\label{table:1}      
\centering          
\begin{tabular}{c  c  c  c  c  c  c}
\hline\hline \\ [-1.5ex]       
ID & Gaia DR3 ID & RA & Dec & T$_{\text{eff}}$ & $\log g$ & $\xi$  \\
NGC~1786- & Gaia DR3 & [J2016] & [J2016] & [K] & [dex] & [km s$^{-1}$] \\ [+0.5ex] 
\hline \\ [-1.5ex]
978 & 4661651610683717248 & 74.7878641 & -67.7285246 & 4310 & 0.83 & 1.40 \\ [0.4ex] 
1321 & 4661649995776473600 & 74.7638489 & -67.7546146 & 4300 & 0.81 & 1.90 \\ [0.4ex] 
1436 & 4661654462541989248 & 74.7555606 & -67.7353347 & 4350 & 0.90 & 1.50 \\ [0.4ex] 
1501 & 4661652946395262336 & 74.7493142 & -67.7514295 & 4250 & 0.73 & 1.90 \\ [0.4ex] 
2310 & 4661651473244758656 & 74.7588569 & -67.7432595 & 4150 & 0.56 & 2.10 \\ [0.4ex] 
2418 & 4661651541942717696 & 74.8215213 & -67.7387519 & 4180 & 0.61 & 1.60 \\ [0.4ex] 
\hline                  
\end{tabular}
%\tablefoot{blabla.}
\end{table*}

\section{Observations} \label{sect_obs}
The sample stars used in this analysis were originally observed by \citet{mucciarelli_09}, who identified Na–O and Mg–Al anticorrelations in NGC~1786. The same stars were reanalyzed by \citet{mucciarelli_21}. In this work we analyzed six out of the seven stars present in the original  sample\footnote{During the new run of observations one star was incorrectly targeted due to a pointing error. This resulted in the observation of the wrong star, and therefore we obtained a final sample of 6 stars.}. All these targets are located in the bright portion of the red giant branch.
The stars were observed with the high-resolution MIKE spectrograph \citep{bernstein_03} at the Magellan/Clay telescope over two runs in November 2023 and October 2024 (Program IDs:CN2023B-37, CN2024B-53; PI: L. Monaco). 
The total integration time on each source ranged from 1800 s up to 4500 s, depending on the target magnitude. We adopted the $0.7" \times 5"$ slit and $2\times 2$ binning, resulting in a resolving power of 28000 in the red arm (4900-9500 $\angstrom$). 
%We made use of the spectra covering the red part of the spectrograph (4900-9500 $\angstrom$) acquired using the $0.7" \times 5"$ slit and $2\times 2$ binning, resulting in a resolving power of 28000. 
The data were reduced with the CarPy MIKE pipeline \citep{kelson_03}. While Fe and Mg abundances for these stars were already present in the literature, these spectra allowed us to derive for the first time K abundances from the K~I resonance line at 7699 $\angstrom$. We checked for each target that K line was not contaminated by telluric lines. This is due to the high radial velocity (RV) of NGC~1786 (264.3 km s$^{-1}$, $\sigma = 5.7$ km s$^{-1}$; \citealt{mucciarelli_09}). The final signal-to-noise ratio per pixel is of $\sim 35$ around 5700 $\angstrom$ and of $\sim 65$ around 7700 $\angstrom$.

%-------------------------------------------------------------------
\section{Chemical analysis} \label{sect_analysis}
We employed for the target stars the atmospheric parameters (T$_{\text{eff}}$, $\log g$, and $\xi$) and their associated errors as derived by \citet{mucciarelli_21}. In particular, T$_{\text{eff}}$ were obtained spectroscopically, by erasing any trend between the abundances and 
the excitation potential of the Fe~I lines, and then corrected to bring them on a photometric-based temperature scale 
with Equation 2 of \citet{mucciarelli_20} in order to remove biases affecting spectroscopic T$_{\text{eff}}$ in metal-poor giant stars 
\citep{frebel13,mucciarelli_20}.
All the relevant information about the observed targets (IDs, Gaia DR3 IDs, coordinates, and the adopted atmospheric parameters) are reported in Table \ref{table:1}. 

Chemical abundances were determined via spectral synthesis. For this purpose, we employed our in-house code \texttt{SALVADOR} (\textit{Alvarez Garay et al., in prep}), which performs a $\chi^2$ minimization between the observed line and a grid of suitable synthetic spectra calculated on the fly using the code \texttt{SYNTHE} \citep{kurucz_05} in which only the abundance of the element under analysis is varied. 
In the calculation of the synthetic spectra we included all the atomic and molecular transitions available in the Kurucz-Castelli database\footnote{\url{https://wwwuser.oats.inaf.it/fiorella.castelli/linelists.html}} and we adopted ATLAS9 model atmospheres. We calculated the total uncertainty associated with the abundances ([Mg/Fe] and [K/Fe]) following the procedure described in \citet{mucciarelli_21} that takes into account the internal errors and the errors arising from the adopted atmospheric parameters. 

The Fe abundances for the target stars were adopted from the comprehensive analysis by \citet{mucciarelli_21}. However, as a sanity check, we performed an independent analysis by measuring Fe abundances from our spectra using a suitable linelist of unblended and unsaturated lines at the MIKE resolution. The results of this check were fully consistent with the values reported by \citet{mucciarelli_21}, therefore we decided to keep the Fe abundances that they derived. Also Mg abundances were already present in the literature, as obtained by \citet{mucciarelli_09}. However, in this case we decided to derive newly Mg since the atmospheric parameters reported by \citet{mucciarelli_09} are slightly different from those present in \citet{mucciarelli_21}\footnote{The newly derived [Mg/Fe] abundance ratios are $\sim 0.02$ dex lower than those presented in \citet{mucciarelli_09}, thus resulting fully consistent with their values.}. We made use of the Mg lines at 5528, 5711, and 8806 $\angstrom$. Corrections for the departure from the Local Thermal Equilibrium (LTE) were from the grids of corrections by \citet{bergemann_17}. We corrected the K abundances derived from the line at 7699 $\angstrom$ for the Non-LTE effects by interpolating into the grids of \citet{reggiani_19}. Finally, as reference solar abundances we adopted those computed by \citet{lodders_10} for the Mg and by \citet{caffau_11} for the K. 
Table \ref{table:2} lists the abundance ratios and the corresponding total uncertainties for all the stars observed in NGC~1786.

%-------------------------------------------------------------------
\section{Results} \label{sect_res}

%-------------
\begin{table}
\caption{Derived elemental abundances for the NGC~1786 stars.}
\label{table:2}
\centering
%\renewcommand{\arraystretch}{1.25}
%\scriptsize
%\setlength{\tabcolsep}{9pt}
\begin{tabular}{c c c c}
\hline\hline \\ [-1.5ex]
ID & [Fe/H] & [Mg/Fe] & [K/Fe] \\
NGC1786- & 7.52 & 7.54 & 5.11  \\ [+0.5ex]
\hline \\ [-1.5ex]
%SUN  &   7.52 & 7.54 & 5.11 \\ [0.4ex]
978  & $-1.67 \pm 0.10$ & $ +0.20 \pm 0.04$ & $+0.29 \pm 0.11$   \\ [0.4ex]
1321 & $-1.75 \pm 0.07$ & $ +0.43 \pm 0.04$ & $+0.15 \pm 0.11$   \\ [0.4ex] 
1436 & $-1.68 \pm 0.10$ & $ +0.37 \pm 0.06$ & $+0.32 \pm 0.12$   \\ [0.4ex] 
1501 & $-1.67 \pm 0.07$ & $ +0.42 \pm 0.05$ & $+0.27 \pm 0.11$   \\ [0.4ex] 
2310 & $-1.72 \pm 0.06$ & $-0.28 \pm 0.06$ & $+0.54 \pm 0.10$   \\ [0.4ex] 
2418 & $-1.68 \pm 0.12$ & $-0.24 \pm 0.05$ & $+0.46 \pm 0.12$   \\ [0.4ex] \hline\hline
\end{tabular}
\vspace{1pt}
\end{table}
%-------------
\begin{figure}
   \centering
   \includegraphics[width=9cm]{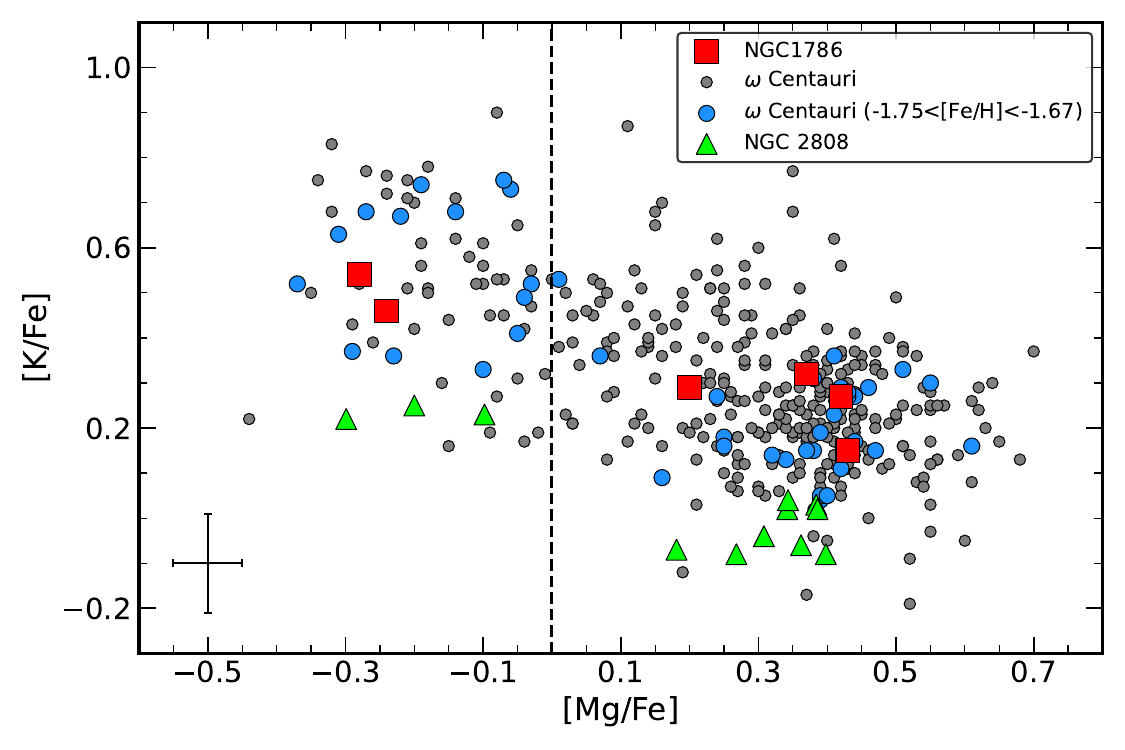}
      \caption{Run of [K/Fe] as a function of [Mg/Fe] for the six stars of NGC~1786 here analyzed (red squares). As a comparison the same trend is plotted for the stars belonging to $\omega$ Centauri (gray and blue circles; \citealt{alvarezgaray_22}) and to NGC~2808 (green triangles; \citealt{mucciarelli_15}). The blue circles represent the stars of $\omega$ Centauri in the same metallicity range of NGC~1786. The vertical dashed line splits the Mg-poor and Mg-rich stars. The error bar on the bottom-left corner represents the typical error associated to the abundance ratios.}
      \label{fig:mgk_anti}
\end{figure}
%------------
%SOMETHING ABOUT THE RV? THAT THEY ARE IN PERFECT AGREENENT WITH THE ONES REPORTED BY MUCCIARELLI+09, INDICATING THAT THE STARS UNDER ANALYSIS ARE ISOLATED AND GOOD STARS?
Figure \ref{fig:mgk_anti} shows the distribution of the [K/Fe] abundance ratios as a function of [Mg/Fe] for the six stars of NGC~1786 analyzed in this study. 
A clear Mg-K anticorrelation emerges to exists among the stars of NGC~1786, despite the small size of the dataset. 
The four Mg-rich stars of NGC~1786 are characterized by $\left\langle \text{[Mg/Fe]} \right\rangle  = +0.36\ (\sigma = 0.09)$ dex and $\left\langle \text{[K/Fe]} \right\rangle  = +0.26\ (\sigma = 0.06)$ dex. On the other hand, the two Mg-poor stars have $\left\langle \text{[Mg/Fe]} \right\rangle  = -0.26\ (\sigma = 0.02)$ dex and $\left\langle \text{[K/Fe]} \right\rangle  = +0.50\ (\sigma = 0.04)$ dex. 
Following the widely accepted nomenclature for GC stars, the four Mg-rich/K-normal stars belong to the 1P, while the two Mg-poor/K-enhanced stars belong to the 2P.
The average [K/Fe] values of the two groups of stars are clearly not compatible each other within the uncertainties. 
A Welch test between the two mean values provides a p-value of 0.02, rejecting the null hypothesis that the two values are compatible each other.

\begin{figure*}
   \centering
   \includegraphics[width=0.9\textwidth]{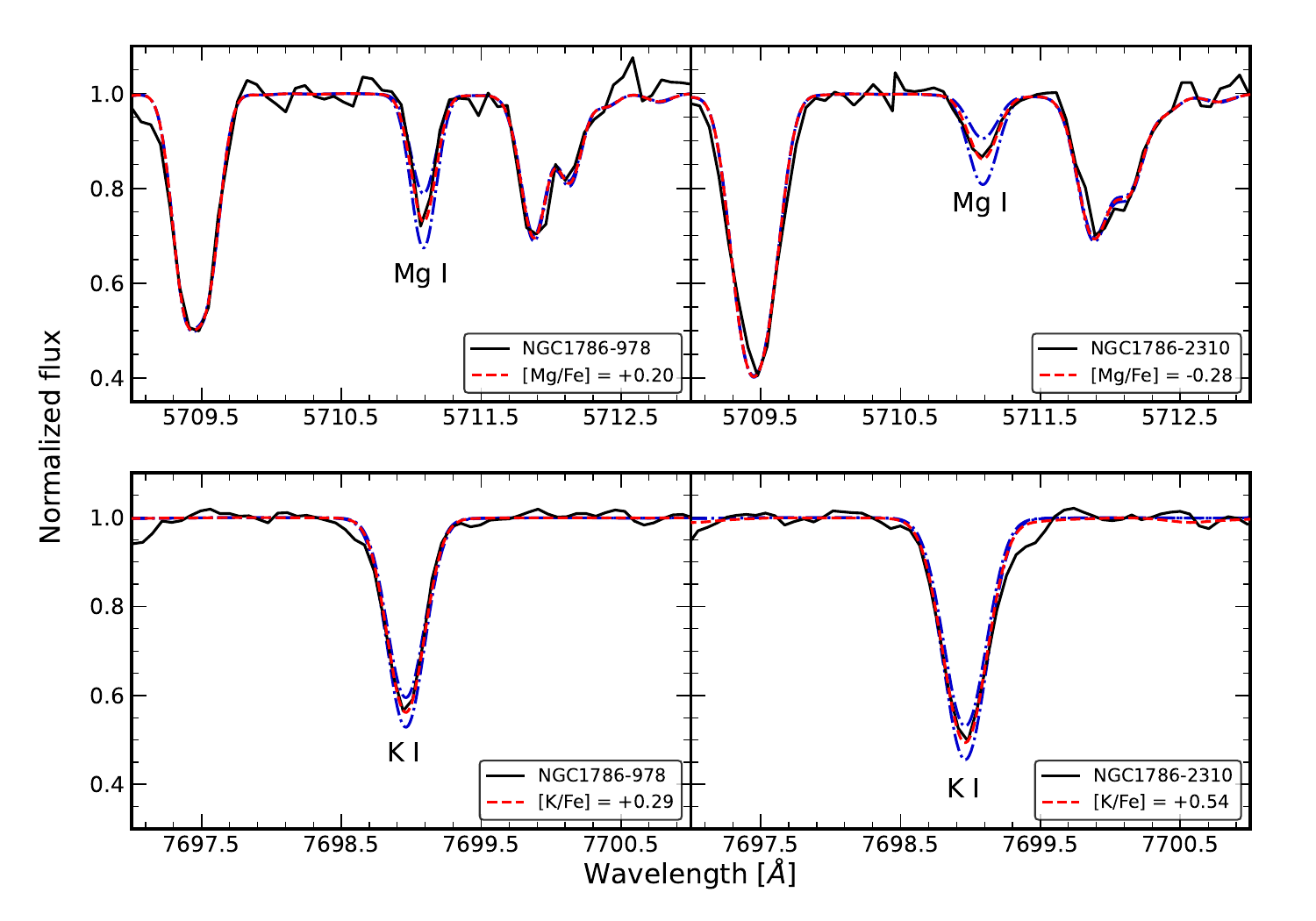}
      \caption{Comparison between the spectra of NGC1786-978 (left panels) and NGC1786-2310 (right panels) around the Mg line at 5711 $\angstrom$ (upper panels) and the K line at 7699 $\angstrom$ (lower panels). The solid black lines represent the observed spectra, the solid red lines the best-fit synthetic spectra, and the dash-dotted blue lines the synthetic spectra with the abundances varied by $\pm 0.2$ dex with respect to the best values, respectively. In the bottom-right corners are reported the name of the stars and the corresponding [Mg/Fe] and [K/Fe] abundance ratios.       
      } %The gray dash-dotted lines mark the position of the Mg and K transitions in both panels. NGC1786-978 (Mg-rich) has a deeper/shallower Mg/K line compared to NGC1786-2310 (Mg-poor). This difference points toward a real difference in the abundances of Mg and K in these stars.}
         \label{fig:mgk_lines}
\end{figure*}

We checked the statistical significance of the observed Mg-K anticorrelation. 
The Spearman correlation test provides a probability that the two abundance ratios are non-correlated 
of 0.5\% (with a $\rho$ coefficient of --0.94).
As an additional support to our finding of the presence of a Mg-K anticorrelation in NGC~1786, in Fig. \ref{fig:mgk_lines} we display
portions of the MIKE spectra around the Mg line at 5711 $\angstrom$ and K line at 7699 $\angstrom$ for two stars of NGC~1786, with superimposed the best-fit synthetic spectra derived from the analysis.  
The figure clearly shows a difference in the lines depth, implying intrinsic differences in the derived Mg and K abundances, given that the stars have the same metallicity and similar atmospheric parameters.. 

While the amplitude of the anticorrelation in NGC~1786 is relatively weak (with variation of $\sim$0.3 dex in [K/Fe] compared to an observed variation of $\sim$0.7 dex in [Mg/Fe]), the observed trend is analogous to those found in NGC~2808 and $\omega$ Centauri, two of the few GC-like systems hosting Mg-poor sub-populations (see Fig. \ref{fig:mgk_anti}).
NGC~2808 is a genuine GC, among the most massive ones \citep[$\rm 7.9\times10^{5}~M_{\odot}$,][]{baum18} and with a higher metallicity than NGC 1786 \citep[\fe $= -1.13$ dex;][]{carretta_15}. 
It shows a Mg-K anticorrelation qualitatively similar to that of NGC~1786 but offset by ~0.15 dex lower in [K/Fe] \citep{mucciarelli_15}. This offset is likely attributable to the different NLTE corrections adopted in the two studies or to the different metallicity regimes of the two clusters. Indeed, NGC~2808 is more metal-rich than NGC~1786 and the the efficiency of the K-enrichment scales with the metallicity, being the MgAl cycle more and more efficient towards the metal-poor regime.
Also noteworthy is the comparison with $\omega$ Centauri, a complex stellar system that, 
in addition to showing a metallicity spread typical of galaxies \citep{johnson10,mesazros20}, also displays all the characteristic features of the self-enrichment processes seen in globular clusters 
\citep{norris95,marino11,mesazros20}, including the Mg-K anticorrelation \citep{alvarezgaray_22}.
Indeed, this GC-like system also exhibits a Mg-K anticorrelation that overlaps with that of NGC 1786, especially when considering only stars of similar metallicity. In fact, when we consider the sub-sample of 42 stars of $\omega$ Centauri in the same metallicity range of NGC~1786 ($-1.75 <$ [Fe/H] $<-1.67$ dex), the stellar group with [Mg/Fe]$>$0.0 dex has $\left\langle \text{[K/Fe]} \right\rangle  = +0.21\ (\sigma = 0.13)$ dex, 
while the Mg-poor stars have $\left\langle \text{[K/Fe]} \right\rangle  = +0.56\ (\sigma = 0.15)$ dex. 
A different case is NGC 2419, which displays a Mg-K anticorrelation that is completely off-scale compared to those defined by $\omega$ Centauri, NGC~2808, and NGC~1786. 
Finally, a different situation arises when comparing NGC~1786 to NGC~4833 and M~54. In both of these GCs, the Mg-K anticorrelation is less extended than in NGC~1786, due to the lack of a sub-population of Mg-poor stars \citep{carretta_21,carretta_22}.

%In particular, from the total sample of $\omega$ Centauri stars \citet{alvarezgaray_22} we have a sub-sample of 42 stars in the same metallicity range of NGC~1786 ($-1.75 <$ [Fe/H] $<-1.67$ dex). Among them, 28 stars belong to the Mg-rich group and have $\left\langle \text{[Mg/Fe]} \right\rangle  = 0.37, (\sigma = 0.11)$ dex and $\left\langle \text{[K/Fe]} \right\rangle  = 0.21, (\sigma = 0.13)$ dex. The remaining 14 stars belong to the Mg-poor group and have $\left\langle \text{[Mg/Fe]} \right\rangle  = -0.17, (\sigma = 0.11)$ dex and $\left\langle \text{[K/Fe]} \right\rangle  = 0.56, (\sigma = 0.15)$ dex. \\

%\textit{This work presents the first clear evidence of the rare Mg-K anticorrelation in an extragalactic GC. This finding strongly suggests that the formation of this extreme chemical signature is independent of the host galaxy environment.} Instead, the Mg-K anticorrelation appears to be a fundamental feature of any old, massive, and metal-poor GC that contains a sub-population of Mg-poor stars. In particular, the two Mg-poor stars (NGC1786-2310 and NGC1786-2418) are enhanced in [K/Fe] by $\sim 0.25$ with respect to the other four Mg-rich stars with normal [K/Fe] content. Following the widely accepted nomenclature for GC  stars, the four Mg-rich/K-normal stars correspond to the 1P, while the two Mg-poor/K-enhanced stars correspond to the 2P.

%-------------
%------------

%-------------------------------------------------------------------
\section{Discussion and Conclusions} \label{sect_disc}

This work presents the first clear evidence of the rare Mg-K anticorrelation in a GC residing in an external galaxy. 
This finding strongly suggests that the formation of this extreme chemical signature is independent of the host galaxy environment. 
Instead, the Mg-K anticorrelation appears to be a ubiquitous feature of any old, massive, and metal-poor GC that contains a sub-population of Mg-poor stars.  
In fact, in NGC~1786, the Mg-poor stars are enhanced in [K/Fe] by $\sim+0.25$ dex with respect to the Mg-rich stars, in perfect agreement with the findings in NGC~2808 and $\omega$ Centauri, both residing in the Milky Way. Once again, NGC~2419 proves to be a unique case among this limited group, with an extension of the Mg-K anticorrelation \citep{mucciarelli_12,cohen_12} that is completely off-scale compared to the other systems where it has been identified. In the AGB scenario, the most common explanation for this peculiar system is that the 2P stars formed directly from the winds of the most massive AGB stars (M $> 6~\mathrm{M}_{\odot}$; \citealt{ventura_12,ventura_18}), with a small or negligible dilution with pristine gas\footnote{With this assumption is possible to explain the presence of stars with very low/high content of Mg/K.}. The model of a 2P formed from gas with a negligible dilution seems to be reinforced also by the fact that \citet{di_criscienzo_15} showed that the Horizontal Branch of NGC~2419 can be explained only by invoking the presence of a He-rich population, with Y $> 0.35$.  Dilution with pristine gas, if any, must have been negligible in this case, otherwise the helium content of these 2P stars would be smaller, as a consequence of mixing of He-rich matter from the AGB winds.

%Models for the formation of such extreme 2P stars were proposed by \citet{d'ercole_08,d'ercole_16,d'antona_16}. 
%t is possible that the peculiar com-bination of low-metallicity, large mass, and the large distance from the main parent Galaxy could explain the observed signatures of this object. 

The discovery of the Mg-K anticorrelation in NGC~1786 provides a new, crucial piece of evidence for understanding the most extreme chemical enrichment processes in GCs. 
The detection of the Mg-K anticorrelation in an LMC cluster strongly suggests that this extreme chemical signature is a universal feature of a specific class of GC, independent of the host galaxy environment. This conclusion echoes previous findings for the Na-O and Mg-Al anticorrelations, which were also confirmed in other LMC GCs \citep{mucciarelli_09} and interpreted as evidence that the self-enrichment process is governed by the cluster's intrinsic properties, primarily mass and metallicity, rather than its galactic environment. 

However, explaining the mechanism producing the Mg-K anticorrelation (coupled with all the other spectroscopic and photometric 
evidence of self-enrichment in GCs) remains a significant challenge for all the proposed models for multiple populations in GCs. 
In fact, K is produced through the reaction $^{36}\text{Ar(p,}\gamma) ^{37}\text{K}(\beta^+) ^{37}\text{Ar(p,}\gamma) ^{38}\text{K}(\beta^+) ^{38}\text{Ar(p,}\gamma) ^{39}\text{K}$. 
This reaction requires very high temperatures (T $> 10^8$ K) to activate proton-capture reactions on Ar nuclei, while simultaneously depleting Mg \citep{ventura_12,iliadis_16,prantzos_17}. 
Based on current theoretical models, the polluter scenario involving Asymptotic Giant Branch (AGB) and super-AGB stars appears to provide the most plausible framework for interpreting our results in NGC~1786, even though the model requires some degree of fine tuning regarding the dilution of the ejected material with pristine gas present in the intra-cluster medium. 
This scenario is, at present, the only one that predicts both the strong Mg depletion and the production of K within the same stellar environment during the hot bottom burning phase \citep{ventura_12,ventura_18}. In contrast, alternative models face significant difficulties while trying to explain the observed overabundance of K. Among the polluter candidates there are fast rotating massive stars \citep{krause_13}, interacting binaries \citep{mink_09}, supermassive stars \citep[M $\sim 1000 \text{M}_{\odot}$][]{denissenkov_hart_14}.
With different degrees of fine tuning and adjustments in the reaction rates these models are able to activate the CNO cycle and the secondary chains in their interiors and reproduce some of the observed anticorrelations (Na-O and Mg-Al) but they fail in the explanation of the Mg-K anticorrelation \citep[see e.g.;][for a discussion]{renzini_15}. 
Also novae \citep{maccarone_12,denissenkov_14} were proposed as possible polluters able to produce K, although they too face significant challenges in reproducing the full chemical pattern observed so far. Indeed, in the model of novae elements such as Na, Al, Si, and S are systematically overproduced to amount much larger than what is observed in GCs. 
However, the study of the nova nucleosynthesis products is not straightforward since the nova outburst is a multi-parameter phenomenon that depends on the mass, temperature, and composition of the exploding dwarf, but also on the accretion rate and composition of the material of the companion star, and many other parameters \citep[see][for a discussion]{jose_17}. Therefore, a thorough investigation of the
large parameter space of novae is needed to better understand this model.

Even though the AGB and super-AGB stars are able reproduce the observed Mg-K anticorrelation in those clusters where this feature is present, the model suffers for some flaws especially regarding the precise physics and timing of the dilution process. However, interestingly in the model for the old and metal-poor Galactic GC NGC~2419 by \citet{ventura_12} they predict that the oxygen abundance must be depleted by a huge factor in the Mg–poor stars. In NGC~1786, the two Mg-poor stars (NGC1786-2310 and NGC1786-2418) are classified as super-O-poor objects \citep{mucciarelli_09}, reinforcing the AGB and super-AGB model as the most valuable for the explanation of multiple populations in this cluster. 

In conclusion, NGC~1786 is an extragalactic GC in which all the chemical anticorrelations linked to the CNO cycle and the secondary chains NeNa and MgAl are clearly visible. Indeed, this system shows evident Na-O, Mg-Al \citep{mucciarelli_09}, and Mg-K anticorrelations. 
These results suggest that this cluster experienced the complete self-enrichment process observed in other Galactic GCs such as NGC~2419, NGC~2808, and $\omega$ Centauri. 
The similarities of NGC~1786 with these Galactic GCs points forward to the idea that there is a global mass/metallicity threshold effect that allows only to the most massive and/or metal-poor GCs to experience the complete phenomenon of multiple populations, independently from the environment in which these systems were born.

Indeed, the link to cluster mass and metallicity is substantiated by comparing the GCs known to host the Mg-K anticorrelation with those that do not. In fact, clusters exhibiting this feature (or alternatively a spread in [K/Fe]), such as NGC~2419, NGC~2808, $\omega$ Centauri, M~54 and NGC~4833 are among the most massive systems \citep{harris_10,baum18}, and with the exception of NGC~2808 are also metal-poor. Conversely, extensive searches in less massive clusters like NGC~6752 and NGC~6809, or in the massive but more metal-rich NGC~104, did not revealed a Mg-K anticorrelation  \citep{mucciarelli_17} nor significant K spread, suggesting that both high mass/low metallicity are necessary conditions to trigger this extreme nucleosynthetic path. The finding of Mg-K anticorrelation in NGC~1786 nicely fits within this framework 
suggesting that the self-enrichment phenomena in clusters formed in the LMC (and in general in other environments) are shaped as in our Galaxy.

\begin{acknowledgements}
Funded by the European Union (ERC-2022-AdG, "StarDance: the non-canonical evolution of stars in clusters", Grant Agreement 101093572, PI: E. Pancino). Views and opinions expressed are however those of the author(s) only and do not necessarily reflect those of the European Union or the European Research Council. Neither the European Union nor the granting authority can be held responsible for them.
A.M. acknowledges support from the project "LEGO – Reconstructing the building blocks of the Galaxy by chemical tagging" (P.I. A. Mucciarelli) granted by the Italian MUR through contract PRIN 2022LLP8TK\_001. 
L.M. gratefully acknowledges support from ANID-FONDECYT Regular Project n. 1251809.

\end{acknowledgements}
\bibliographystyle{aa} % style aa.bst
\bibliography{biblio} % your references Yourfile.bib

\begin{thebibliography}{46}
\expandafter\ifx\csname natexlab\endcsname\relax\def\natexlab#1{#1}\fi

\bibitem[{{Alvarez Garay} {et~al.}(2022){Alvarez Garay}, {Mucciarelli}, {Lardo}, {Bellazzini}, \& {Merle}}]{alvarezgaray_22}
{Alvarez Garay}, D.~A., {Mucciarelli}, A., {Lardo}, C., {Bellazzini}, M., \& {Merle}, T. 2022, \apjl, 928, L11

\bibitem[{{Bastian} \& {Lardo}(2018)}]{bastian_18}
{Bastian}, N. \& {Lardo}, C. 2018, \araa, 56, 83

\bibitem[{{Baumgardt} \& {Hilker}(2018)}]{baum18}
{Baumgardt}, H. \& {Hilker}, M. 2018, \mnras, 478, 1520

\bibitem[{{Bergemann} {et~al.}(2017){Bergemann}, {Collet}, {Amarsi}, {Kovalev}, {Ruchti}, \& {Magic}}]{bergemann_17}
{Bergemann}, M., {Collet}, R., {Amarsi}, A.~M., {et~al.} 2017, \apj, 847, 15

\bibitem[{{Bernstein} {et~al.}(2003){Bernstein}, {Shectman}, {Gunnels}, {Mochnacki}, \& {Athey}}]{bernstein_03}
{Bernstein}, R., {Shectman}, S.~A., {Gunnels}, S.~M., {Mochnacki}, S., \& {Athey}, A.~E. 2003, in Society of Photo-Optical Instrumentation Engineers (SPIE) Conference Series, Vol. 4841, Instrument Design and Performance for Optical/Infrared Ground-based Telescopes, ed. M.~{Iye} \& A.~F.~M. {Moorwood}, 1694--1704

\bibitem[{{Brocato} {et~al.}(1996){Brocato}, {Castellani}, {Ferraro}, {Piersimoni}, \& {Testa}}]{brocato96}
{Brocato}, E., {Castellani}, V., {Ferraro}, F.~R., {Piersimoni}, A.~M., \& {Testa}, V. 1996, \mnras, 282, 614

\bibitem[{{Caffau} {et~al.}(2011){Caffau}, {Ludwig}, {Steffen}, {Freytag}, \& {Bonifacio}}]{caffau_11}
{Caffau}, E., {Ludwig}, H.~G., {Steffen}, M., {Freytag}, B., \& {Bonifacio}, P. 2011, \solphys, 268, 255

\bibitem[{{Carretta}(2015)}]{carretta_15}
{Carretta}, E. 2015, \apj, 810, 148

\bibitem[{{Carretta}(2021)}]{carretta_21}
{Carretta}, E. 2021, \aap, 649, A154

\bibitem[{{Carretta}(2022)}]{carretta_22}
{Carretta}, E. 2022, \aap, 666, A177

\bibitem[{{Cohen} \& {Kirby}(2012)}]{cohen_12}
{Cohen}, J.~G. \& {Kirby}, E.~N. 2012, \apj, 760, 86

\bibitem[{{de Mink} {et~al.}(2009){de Mink}, {Pols}, {Langer}, \& {Izzard}}]{mink_09}
{de Mink}, S.~E., {Pols}, O.~R., {Langer}, N., \& {Izzard}, R.~G. 2009, \aap, 507, L1

\bibitem[{{Denissenkov} \& {Hartwick}(2014)}]{denissenkov_hart_14}
{Denissenkov}, P.~A. \& {Hartwick}, F.~D.~A. 2014, \mnras, 437, L21

\bibitem[{{Denissenkov} {et~al.}(2014){Denissenkov}, {Truran}, {Pignatari}, {Trappitsch}, {Ritter}, {Herwig}, {Battino}, {Setoodehnia}, \& {Paxton}}]{denissenkov_14}
{Denissenkov}, P.~A., {Truran}, J.~W., {Pignatari}, M., {et~al.} 2014, \mnras, 442, 2058

\bibitem[{{Di Criscienzo} {et~al.}(2015){Di Criscienzo}, {Tailo}, {Milone}, {D'Antona}, {Ventura}, {Dotter}, \& {Brocato}}]{di_criscienzo_15}
{Di Criscienzo}, M., {Tailo}, M., {Milone}, A.~P., {et~al.} 2015, \mnras, 446, 1469

\bibitem[{{Frebel} {et~al.}(2013){Frebel}, {Casey}, {Jacobson}, \& {Yu}}]{frebel13}
{Frebel}, A., {Casey}, A.~R., {Jacobson}, H.~R., \& {Yu}, Q. 2013, \apj, 769, 57

\bibitem[{{Gaia Collaboration} {et~al.}(2016){Gaia Collaboration}, {Prusti}, {de Bruijne}, {Brown}, {Vallenari}, {Babusiaux}, {et~al.}}]{gaia_16}
{Gaia Collaboration}, {Prusti}, T., {de Bruijne}, J.~H.~J., {et~al.} 2016, \aap, 595, A1

\bibitem[{{Gaia Collaboration} {et~al.}(2023){Gaia Collaboration}, {Vallenari}, {Brown}, {Prusti}, {de Bruijne}, {Arenou}, {et~al.}}]{gaia_23}
{Gaia Collaboration}, {Vallenari}, A., {Brown}, A.~G.~A., {et~al.} 2023, \aap, 674, A1

\bibitem[{{Gratton} {et~al.}(2019){Gratton}, {Bragaglia}, {Carretta}, {D'Orazi}, {Lucatello}, \& {Sollima}}]{gratton_19}
{Gratton}, R., {Bragaglia}, A., {Carretta}, E., {et~al.} 2019, \aapr, 27, 8

\bibitem[{{Harris}(2010)}]{harris_10}
{Harris}, W.~E. 2010, arXiv e-prints, arXiv:1012.3224

\bibitem[{{Iliadis} {et~al.}(2016){Iliadis}, {Karakas}, {Prantzos}, {Lattanzio}, \& {Doherty}}]{iliadis_16}
{Iliadis}, C., {Karakas}, A.~I., {Prantzos}, N., {Lattanzio}, J.~C., \& {Doherty}, C.~L. 2016, \apj, 818, 98

\bibitem[{{Johnson} \& {Pilachowski}(2010)}]{johnson10}
{Johnson}, C.~I. \& {Pilachowski}, C.~A. 2010, \apj, 722, 1373

\bibitem[{{Jos{\'e}}(2017)}]{jose_17}
{Jos{\'e}}, J. 2017, in 14th International Symposium on Nuclei in the Cosmos (NIC2016), ed. S.~{Kubono}, T.~{Kajino}, S.~{Nishimura}, T.~{Isobe}, S.~{Nagataki}, T.~{Shima}, \& Y.~{Takeda}, 010501

\bibitem[{{Kelson}(2003)}]{kelson_03}
{Kelson}, D.~D. 2003, \pasp, 115, 688

\bibitem[{{Krause} {et~al.}(2013){Krause}, {Charbonnel}, {Decressin}, {Meynet}, \& {Prantzos}}]{krause_13}
{Krause}, M., {Charbonnel}, C., {Decressin}, T., {Meynet}, G., \& {Prantzos}, N. 2013, \aap, 552, A121

\bibitem[{{Kurucz}(2005)}]{kurucz_05}
{Kurucz}, R.~L. 2005, Memorie della Societa Astronomica Italiana Supplementi, 8, 14

\bibitem[{{Langer} {et~al.}(1993){Langer}, {Hoffman}, \& {Sneden}}]{langer_93}
{Langer}, G.~E., {Hoffman}, R., \& {Sneden}, C. 1993, \pasp, 105, 301

\bibitem[{{Lodders}(2010)}]{lodders_10}
{Lodders}, K. 2010, in Astrophysics and Space Science Proceedings, Vol.~16, Principles and Perspectives in Cosmochemistry, ed. A.~{Goswami} \& B.~E. {Reddy}, 379

\bibitem[{{Maccarone} \& {Zurek}(2012)}]{maccarone_12}
{Maccarone}, T.~J. \& {Zurek}, D.~R. 2012, \mnras, 423, 2

\bibitem[{{Mackey} \& {Gilmore}(2003)}]{mackey2003}
{Mackey}, A.~D. \& {Gilmore}, G.~F. 2003, \mnras, 338, 85

\bibitem[{{Marino} {et~al.}(2011){Marino}, {Milone}, {Piotto}, {Villanova}, {Gratton}, {D'Antona}, {Anderson}, {Bedin}, {Bellini}, {Cassisi}, {Geisler}, {Renzini}, \& {Zoccali}}]{marino11}
{Marino}, A.~F., {Milone}, A.~P., {Piotto}, G., {et~al.} 2011, \apj, 731, 64

\bibitem[{{M{\'e}sz{\'a}ros} {et~al.}(2020){M{\'e}sz{\'a}ros}, {Masseron}, {Garc{\'\i}a-Hern{\'a}ndez}, {Allende Prieto}, {Beers}, {Bizyaev}, {Chojnowski}, {Cohen}, {Cunha}, {Dell'Agli}, {Ebelke}, {Fern{\'a}ndez-Trincado}, {Frinchaboy}, {Geisler}, {Hasselquist}, {Hearty}, {Holtzman}, {Johnson}, {Lane}, {Lacerna}, {Longa-Pe{\~n}a}, {Majewski}, {Martell}, {Minniti}, {Nataf}, {Nidever}, {Pan}, {Schiavon}, {Shetrone}, {Smith}, {Sobeck}, {Stringfellow}, {Szigeti}, {Tang}, {Wilson}, \& {Zamora}}]{mesazros20}
{M{\'e}sz{\'a}ros}, S., {Masseron}, T., {Garc{\'\i}a-Hern{\'a}ndez}, D.~A., {et~al.} 2020, \mnras, 492, 1641

\bibitem[{{Milone} \& {Marino}(2022)}]{milone_22}
{Milone}, A.~P. \& {Marino}, A.~F. 2022, Universe, 8, 359

\bibitem[{{Mucciarelli} {et~al.}(2012){Mucciarelli}, {Bellazzini}, {Ibata}, {Merle}, {Chapman}, {Dalessandro}, \& {Sollima}}]{mucciarelli_12}
{Mucciarelli}, A., {Bellazzini}, M., {Ibata}, R., {et~al.} 2012, \mnras, 426, 2889

\bibitem[{{Mucciarelli} {et~al.}(2015){Mucciarelli}, {Bellazzini}, {Merle}, {Plez}, {Dalessandro}, \& {Ibata}}]{mucciarelli_15}
{Mucciarelli}, A., {Bellazzini}, M., {Merle}, T., {et~al.} 2015, \apj, 801, 68

\bibitem[{{Mucciarelli} \& {Bonifacio}(2020)}]{mucciarelli_20}
{Mucciarelli}, A. \& {Bonifacio}, P. 2020, \aap, 640, A87

\bibitem[{{Mucciarelli} {et~al.}(2021){Mucciarelli}, {Massari}, {Minelli}, {Romano}, {Bellazzini}, {Ferraro}, {Matteucci}, \& {Origlia}}]{mucciarelli_21}
{Mucciarelli}, A., {Massari}, D., {Minelli}, A., {et~al.} 2021, Nature Astronomy, 5, 1247

\bibitem[{{Mucciarelli} {et~al.}(2017){Mucciarelli}, {Merle}, \& {Bellazzini}}]{mucciarelli_17}
{Mucciarelli}, A., {Merle}, T., \& {Bellazzini}, M. 2017, \aap, 600, A104

\bibitem[{{Mucciarelli} {et~al.}(2009){Mucciarelli}, {Origlia}, {Ferraro}, \& {Pancino}}]{mucciarelli_09}
{Mucciarelli}, A., {Origlia}, L., {Ferraro}, F.~R., \& {Pancino}, E. 2009, \apjl, 695, L134

\bibitem[{{Norris} \& {Da Costa}(1995)}]{norris95}
{Norris}, J.~E. \& {Da Costa}, G.~S. 1995, \apj, 447, 680

\bibitem[{{Prantzos} {et~al.}(2007){Prantzos}, {Charbonnel}, \& {Iliadis}}]{prantzos_07}
{Prantzos}, N., {Charbonnel}, C., \& {Iliadis}, C. 2007, \aap, 470, 179

\bibitem[{{Prantzos} {et~al.}(2017){Prantzos}, {Charbonnel}, \& {Iliadis}}]{prantzos_17}
{Prantzos}, N., {Charbonnel}, C., \& {Iliadis}, C. 2017, \aap, 608, A28

\bibitem[{{Reggiani} {et~al.}(2019){Reggiani}, {Amarsi}, {Lind}, {Barklem}, {Zatsarinny}, {Bartschat}, {Fursa}, {Bray}, {Spina}, \& {Mel{\'e}ndez}}]{reggiani_19}
{Reggiani}, H., {Amarsi}, A.~M., {Lind}, K., {et~al.} 2019, \aap, 627, A177

\bibitem[{{Renzini} {et~al.}(2015){Renzini}, {D'Antona}, {Cassisi}, {King}, {Milone}, {Ventura}, {Anderson}, {Bedin}, {Bellini}, {Brown}, {Piotto}, {van der Marel}, {Barbuy}, {Dalessandro}, {Hidalgo}, {Marino}, {Ortolani}, {Salaris}, \& {Sarajedini}}]{renzini_15}
{Renzini}, A., {D'Antona}, F., {Cassisi}, S., {et~al.} 2015, \mnras, 454, 4197

\bibitem[{{Ventura} {et~al.}(2012){Ventura}, {D'Antona}, {Di Criscienzo}, {Carini}, {D'Ercole}, \& {vesperini}}]{ventura_12}
{Ventura}, P., {D'Antona}, F., {Di Criscienzo}, M., {et~al.} 2012, \apjl, 761, L30

\bibitem[{{Ventura} {et~al.}(2018){Ventura}, {D'Antona}, {Imbriani}, {Di Criscienzo}, {Dell'Agli}, \& {Tailo}}]{ventura_18}
{Ventura}, P., {D'Antona}, F., {Imbriani}, G., {et~al.} 2018, \mnras, 477, 438

\end{thebibliography}

\end{document}